\documentclass[12pt,a4paper]{article}
\usepackage{epsfig,graphics,amsmath,amssymb,here}
\usepackage{psfrag}

\makeatletter
\newif\if@preliminary
\@preliminaryfalse
\def\preliminary{\@preliminaryfalse}
\def\preprintno#1{\def\@preprintno{#1}}
\def\address#1{\def\@address{#1}}

\def\abstract#1{\def\@abstract{#1}}
\renewcommand\abstractname{ABSTRACT}
\newlength\preprintnoskip
\newlength\abstractwidth
\@titlepagetrue
\renewcommand\maketitle{\begin{titlepage}%
  \let\footnotesize\small
  \hfill\parbox{\preprintnoskip}{%
  \begin{flushright}\@preprintno\end{flushright}}\hspace*{1cm}
  \vskip 60\p@
  \begin{center}%
    {\Large\bf\boldmath \@title \par}\vskip 1cm%
    {\sc\@author \par}\vskip 3mm%
    {\@address \par}%
    \if@preliminary
      \vskip 2cm {\large\sf PRELIMINARY DRAFT \par \@date}%
    \fi
  \end{center}\par
  \@thanks
  \vfill
  \begin{center}%
    \parbox{\abstractwidth}{\centerline{\abstractname}%
    \vskip 3mm%
    \@abstract}
  \end{center}
  \end{titlepage}%
  \setcounter{footnote}{0}%
  \let\thanks\relax\let\maketitle\relax
  \gdef\@thanks{}\gdef\@author{}\gdef\@address{}%
  \gdef\@title{}\gdef\@abstract{}\gdef\@preprintno{}
}%
\def\@citex[#1]#2{\if@filesw\immediate\write\@auxout{\string\citation{#2}}\fi
  \def\@citea{}\@cite{\@for\@citeb:=#2\do
    {\@citea\def\@citea{,\penalty\@m}\@ifundefined
       {b@\@citeb}{{\bf ?}\@warning
       {Citation `\@citeb' on page \thepage \space undefined}}%
\hbox{\csname b@\@citeb\endcsname}}}{#1}}
\def\citerange{\@ifnextchar [{\@tempswatrue\@citexr}{\@tempswafalse\@citexr[]}}
\def\@citexr[#1]#2{\if@filesw\immediate\write\@auxout{\string\citation{#2}}\fi
  \def\@citea{}\@cite{\@for\@citeb:=#2\do
    {\@citea\def\@citea{--\penalty\@m}\@ifundefined
       {b@\@citeb}{{\bf ?}\@warning
       {Citation `\@citeb' on page \thepage \space undefined}}%
\hbox{\csname b@\@citeb\endcsname}}}{#1}}
\long\def\@makecaption#1#2{%
  \vskip\abovecaptionskip
  \sbox\@tempboxa{#1: \emph{#2}}%
  \ifdim \wd\@tempboxa >\hsize
    #1: \emph{#2}\par
  \else
    \hbox to\hsize{\hfil\box\@tempboxa\hfil}%
  \fi
  \vskip\belowcaptionskip}
\def\fmslash{\@ifnextchar[{\fmsl@sh}{\fmsl@sh[0mu]}}
\def\fmsl@sh[#1]#2{%
  \mathchoice
    {\@fmsl@sh\displaystyle{#1}{#2}}%
    {\@fmsl@sh\textstyle{#1}{#2}}%
    {\@fmsl@sh\scriptstyle{#1}{#2}}%
    {\@fmsl@sh\scriptscriptstyle{#1}{#2}}}
\def\@fmsl@sh#1#2#3{\m@th\ooalign{$\hfil#1\mkern#2/\hfil$\crcr$#1#3$}}
\makeatother

\def\fmfL(#1,#2,#3)#4{\put(#1,#2){\makebox(0,0)[#3]{#4}}}

\newcommand\OBS{\mbox{$\cal{O}$}}
\newcommand\ZJ{\mbox{Z $\rightarrow$ 4 jets}}
\newcommand\ZTJ{\mbox{Z $\rightarrow$ 3 jets}}

\newcommand\bb{\mbox{$\bar{b}$}}

\newcommand\ycut{\mbox{$y_{cut}$}}
\newcommand\hvb{\mbox{$\widehat{h}_{Vb}$}}
\newcommand\hab{\mbox{$\widehat{h}_{Ab}$}}
\newcommand\hb{\mbox{$\widehat{h}_{b}$}}
\newcommand\hbn{\mbox{$\widetilde{h}_{b}$}}
\newcommand\dhb{\mbox{$\delta\widehat{h}_{b}$}}
\newcommand\dhbn{\mbox{$\delta\widetilde{h}_{b}$}}
\def\ltap{\raisebox{-.4ex}{\rlap{$\sim$}} \raisebox{.4ex}{$<$}}
\def\gtap{\raisebox{-.4ex}{\rlap{$\sim$}} \raisebox{.4ex}{$>$}}
\begin{document}
%\shortletter        % subdivided in paragraphs instead of sections
\preliminary        % mark on title page
\baselineskip20pt   % stretch linespacing in main text
%%%%%%%%%%%%%%%%%%%%%%%%%%%%%%%%%%%%%%%%%%%%%%%%%%%%%%%%%%%%%%%%%%%%%%%%
\renewcommand{\thefootnote}{\fnsymbol{footnote}}

\begin{flushright}
%\vspace*{-0.3cm}
DESY 03--101\\
%HD--THEP 03--18 \\
LC--TH--2003-097\\
hep-ph/0308198\\[0.3cm]
\end{flushright}

%\preprintno{HD--THEP 03--nn\\hep-ph/0303xxx\\[0.5\baselineskip] March 2003}
\begin{center}
{\Large \bf CP Violation in the 3 Jet and 4 Jet Decays of the \\
%           =================================================
             Z Boson at GigaZ
%           ============================================
} \\[1.0cm]

{\sc O.~Nachtmann$^1$\footnote{Email: O.Nachtmann@thphys.uni-heidelberg.de} and C.~Schwanenberger$^{2}$\footnote{Email: schwanen@mail.desy.de}}\\[1.0cm]
{\it $^1$ Institut f\"ur Theoretische Physik, Universit\"at Heidelberg,
 Philosophenweg 16,\\
 D--69120 Heidelberg, Germany\\
 $^2$ Deutsches Elektronen-Synchrotron DESY, Notkestr. 85, D--22603 Hamburg,
 Germany} \\[1.0cm]
\end{center}

%\footnote{email:}
%\author{%
% O.~Nachtmann\email{{\rm Email: }O.Nachtmann}{thphys.uni-heidelberg.de}
% and C.~Schwanenberger\email{{\rm Email: }C.Schwanenberger}{thphys.uni-heidelberg.de}
% }
%\address{%
% Institut f\"ur Theoretische Physik, Universit\"at Heidelberg,
% Philosophenweg 16\\
% D--69120 Heidelberg, Germany
% }

\begin{abstract}
\noindent
\begin{center}
{\bf Abstract}
\end{center}
 We review CP-violating effects in Z $\rightarrow$ 3 jet and  Z $\rightarrow$ 4 jet decays, assuming the presence of CP-violating effective 
$Z b \bar{b} G$ and $Z b \bar{b} G G$ couplings. Longitudinal beam
polarization is 
included in the studies. We propose a direct 
search for such CP-violating couplings by using various
CP-odd observables. The data of a future linear collider running at the
Z-resonance in the so-called GigaZ option should give  
significant information on the couplings. Finally we show that stringent
bounds on the mass of excited $b$ quarks can be derived if appropriate
couplings are of a size characteristic of a strong interaction.
\end{abstract}

\renewcommand{\thefootnote}{\arabic{footnote}}
\setcounter{footnote}{0}

\section{Introduction}
%%%%%%%%%%%%%%%%%%%%%%%%%%%%%%%%%%%%%%%%%%%%%%%%%%%%%%%%%%%%%%%%%%%%%%%%
%%%%%%%%%%%%%%%%%%%%%%%%%%%%%%%%%%%%%%%%%%%%%%%%%%%%%%%%%%%%%%%%%%%%%%%%

One of the most promising projects in todays high energy physics is an
electron-positron linear collider, for example TESLA \cite{tesla}.
At such a linear collider one should be able to polarize the electrons with
the same technology as at the SLC to up to $80\%$. At TESLA it
should also be possible to run with positrons polarized up to $60\%$
\cite{tesla}. With a luminosity of ${\cal L} \simeq 5 \cdot 10^{33} 
{\rm cm}^{-2} 
{\rm s}^{-1}$ at energies close to the peak of the Z-resonance TESLA could
produce $10^9$ Z-bosons in about 70 days of running \cite{tesla}.
In this scenario, often referred to as GigaZ, the measurements already
performed at the electron-positron collider experiments at LEP and SLC could
be redone with increased precision.

An interesting topic is the test of the CP symmetry in
Z decays. There is already a number of theoretical 
(\cite{othertheo1}-\cite{4jetg} and references therein) and experimental
\cite{alephtautau92}-\cite{thilo}
studies of this subject. In the present paper we will study a
flavor-diagonal Z decay where
CP-violating effects within the
Standard Model (SM) are estimated to be very small \cite{zdecay}. Thus,
looking for CP violation in such Z decays
means looking for new physics beyond the SM. 

%In this paper we deal with the
%question: {\em Does CP violation beyond the SM exist in decays Z
%$\rightarrow$ 4 jets?}
%
For a model-independent systematic analysis of CP violation in Z decays
we use the effective Lagrangian approach as described in
\cite{zdecay,xsec}. Of particular interest are Z decays involving heavy
leptons or quarks. Thus, the process Z $\rightarrow b \bar{b} G$, which is
sensitive to effective CP-violating couplings in the $Z b \bar{b} G$ vertex,
has been analysed theoretically in \cite{width,hab} and experimentally in
\cite{aleph}. No significant deviation from the SM has
been found. 

If CP-violating couplings are introduced in the $Z b \bar{b}
G$ vertex, they will, because of gauge invariance of QCD, appear in the $Z
b \bar{b} G G$ vertex as well. But the $Z b \bar{b} G G$ vertex could in
principle contain new coupling parameters. The analysis of the 4 jet decays of the Z boson
involving $b$ quarks looks into
both, 4- and 5-point vertices. This has been investigated theoretically in
\cite{4jet} and experimentally in 
\cite{thilo}. Also in this case no significant deviation from the SM has
been found. 

In this paper we review the results of the calculations of the processes
\ZTJ\ and \mbox{Z $\rightarrow$ 4 jets} including CP-violating couplings, with at
least two of 
the jets originating from a $b$ or $\bar{b}$ quark, for the GigaZ scenario
assuming longitudinal beam polarization for electrons and positrons.
All details of the calculation for unpolarized $e^+$, $e^-$ beams can be
found in \cite{width,hab,4jet,dok}.

Finally we make an estimate for models with excited quarks and show that
one can obtain stringent bounds on their mass. This, however, requires the
introduction of a new type of strong interactions for quarks.

In chapter 2 we briefly review the theoretical framework of our
computations. Models with excited quarks are discussed in chapter 3. Next,
in chapter 4, we define CP-odd tensor and 
vector observables and calculate their sensitivities to anomalous
couplings. Achievable bounds on the mass of excited quarks are presented.
Our conclusions can be found in chapter 5.

%%%%%%%%%%%%%%%%%%%%%%%%%%%%%%%%%%%%%%%%%%%%%%%%%%%%%%%%%%%%%%%%%%%%%%%%
%%%%%%%%%%%%%%%%%%%%%%%%%%%%%%%%%%%%%%%%%%%%%%%%%%%%%%%%%%%%%%%%%%%%%%%%
\section{Effective Lagrangian Approach}
%%%%%%%%%%%%%%%%%%%%%%%%%%%%%%%%%%%%%%%%%%%%%%%%%%%%%%%%%%%%%%%%%%%%%%%%
%%%%%%%%%%%%%%%%%%%%%%%%%%%%%%%%%%%%%%%%%%%%%%%%%%%%%%%%%%%%%%%%%%%%%%%%

For a model independent study of CP violation in 3 jet and 4 jet
decays of the Z boson we use the effective Lagrangian approach as explained
in \cite{zdecay}. We could add to the SM Lagrangian ${\cal L}_{SM}$ a
CP-violating term ${\cal L}_{CP}$
containing all CP-odd local operators with a mass dimension $d \leq
6$ (\underline{after} electroweak symmetry breaking)
that can be constructed with SM fields. 
However, it turns out that quite a number of such coupling terms can
contribute to the reactions analysed here. To keep the
analysis manageable we restrict ourselves to coupling terms involving the Z
and the $b$ quarks and in addition any number of gluons. Then
the effective CP-violating
Lagrangian with $d \leq 6$ relevant to our analysis is:
\begin{eqnarray}
  \label{lcp} \nonumber
%  \lefteqn{ {\cal L}_{CP}(x) = } \\ \nonumber
{ {\cal L}_{CP}(x) = }
  & - & \frac{i}{2} \widetilde{d}_b \: \bar{b}(x)\: \sigma^{\mu\nu}\: \gamma_5\:
  b(x)\; [\partial_{\mu}\: Z_{\nu}(x) - \partial_{\nu}\: Z_{\mu}(x)] \\
  & + & [\; h_{Vb}\: \bar{b}(x)\: T^a\: \gamma^{\nu}\:
  b(x) +  h_{Ab}\: \bar{b}(x)\: T^a\: \gamma^{\nu}\: \gamma_5\: b(x)\; ]\;
  Z^{\mu}(x)\: 
  G^a_{\mu\nu}(x)\;\;,
\end{eqnarray}
where $b(x)$ denotes the $b$ quark field, $Z^{\mu}(x)$ and $G^a_{\mu\nu}(x)$
represent the field of
the Z boson and the field strength tensor of the gluon, respectively, and
$T^a=\lambda^a/2$ are the 
generators of $SU(3)_C$ \cite{buch}. 
In (\ref{lcp}) $\widetilde{d}_b$ is the weak dipole moment and  $h_{Vb}$,
$h_{Ab}$ are CP-violating
vector and axial vector chirality conserving coupling constants.  
As effective coupling constants in ${\cal L}_{CP}$ the parameters
$\widetilde{d}_b$, $h_{Vb}$, $h_{Ab}$ are real. They are related to
form factors of vertices but should not be confused with the latter
(see e.~g. \cite{remarks}). 

The
Lagrangian ${\cal L}_{CP}$ is required to be invariant under the
electromagnetic and strong gauge group $U(1)_{e.m.} \times SU(3)_C$. We
do not require explicit gauge invariance under the complete $SU(2) \times
U(1)_Y$ 
group of the electroweak interaction, since we consider the theory after
electroweak symmetry breaking. The couplings of (\ref{lcp}) can, however,
always be generated from higher dimensional $SU(2) \times U(1)_Y$ invariant
couplings involving suitable Higgs fields, see~\cite{bernnach2}. The
coupling constants in (\ref{lcp}) are then proportional to powers of the
Higgs vacuum expectation value. Also, 
in theories --- which we do not want to exclude a priori from our
discussion --- where $Z$ and $W^\pm$ are
composite objects the $SU(2) \times U(1)_Y$ group has not necessarily a
fundamental meaning (see the review 
\cite{Harari:1982xy}). Even if compositeness is not favoured by the LEP
data \cite{pdg,radii,tesla} this option for going beyond the SM should
certainly be 
investigated again at the GigaZ factory.

Information on the spin of the final state partons is hardly available
experimentally. Thus, we consider as 
observables only the parton's energies and momenta. Then, effects linear in the
dipole form factor $\widetilde{d}_b$ are suppressed by
powers of $m_b/m_Z$. So angular
correlations of the jets in \ZTJ\ and Z $\rightarrow$ 4 jets 
are only sensitive to the couplings $h_{Vb}$ and
$h_{Ab}$.

%%%%%%%%%%%%%%%%%%begin%%figure%%%%%%%%%%%%%%%%%%%%%%%%%%%%%%%%%%%%%%%
\begin{figure}[ht]
  \begin{center} \epsfig{file=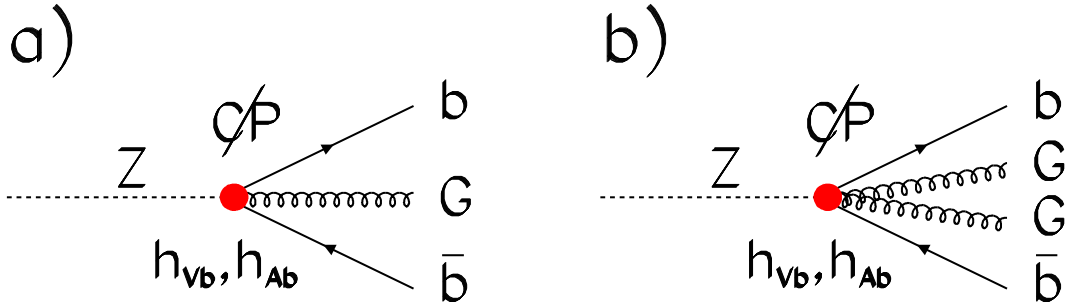,width=\hsize}
    \parbox{0.9\textwidth}
{\caption{\it{
The CP-violating vertices.
    }}\label{fig:vert}
}  \end{center}
\end{figure}
%%%%%%%%%%%%%%%%%%%%%%%%%%%%%%end%%figure%%%%%%%%%%%%%%%%%%%%%%%%%%%%%

The corresponding vertices following from ${\cal L}_{CP}$ are shown 
in figure~\ref{fig:vert}. Because the non-abelian field strength tensor has
a term 
quadratic in the gluon fields the $Z b \bar{b} G$- and $Z b \bar{b} G
G$-vertices are related.

We define dimensionless coupling constants $\widehat{h}_{Vb,Ab}$ using the Z
mass as the scale parameter by
\begin{equation}
\label{hhat}
        h_{Vb,Ab} = \frac{e\: g_s}{\sin \vartheta_W\: \cos \vartheta_W \:m_Z^2}\; \widehat{h}_{Vb,Ab}\;,
\end{equation}
where $e=\sqrt{4\pi \, \alpha}$, $g_s = \sqrt{4\pi \, \alpha_s}$ and
$\vartheta_W$ is the weak mixing angle.
For numerical calculations we set $m_Z=91.187\:$GeV, $\sin^2 \vartheta_W =
0.2236$ and the fine structure constant and $\alpha_s$ at the Z mass
to $\alpha =1/128.9$ and $\alpha_s = 0.118$ \cite{pdg}. Our calculations are carried out in
leading order of the 
CP-violating couplings of
${\cal L}_{CP}$ and the SM couplings. A non-vanishing $b$ quark mass of
$4.5\:$GeV is included \footnote{We use here the pole mass value for the $b$
  quark. In our leading order calculation we could as well use the running
  $b$ mass 
  at $m_Z$: $m_b(m_Z) \simeq 3$ GeV \cite{mbatmz}. This would result only in
  minimal changes in our correlations.}
; masses of $u$, $d$, $s$, $c$ quarks are neglected.

%%%%%%%%%%%%%%%%%%%%%%%%%%%%%%%%%%%%%%%%%%%%%%%%%%%%%%%%%%%%%%%%%%%%%%%%
%%%%%%%%%%%%%%%%%%%%%%%%%%%%%%%%%%%%%%%%%%%%%%%%%%%%%%%%%%%%%%%%%%%%%%%%
\section{Models with excited quarks}
\label{sec:excq}
%%%%%%%%%%%%%%%%%%%%%%%%%%%%%%%%%%%%%%%%%%%%%%%%%%%%%%%%%%%%%%%%%%%%%%%%
%%%%%%%%%%%%%%%%%%%%%%%%%%%%%%%%%%%%%%%%%%%%%%%%%%%%%%%%%%%%%%%%%%%%%%%%

%%%%%%%%%%%%%%%%%%begin%%figure%%%%%%%%%%%%%%%%%%%%%%%%%%%%%%%%%%%%%%%
\begin{figure}[th]
  \begin{center} \epsfig{bbllx=0bp,bblly=130bp,bburx=840bp,bbury=640bp,width=0.6
\hsize,file=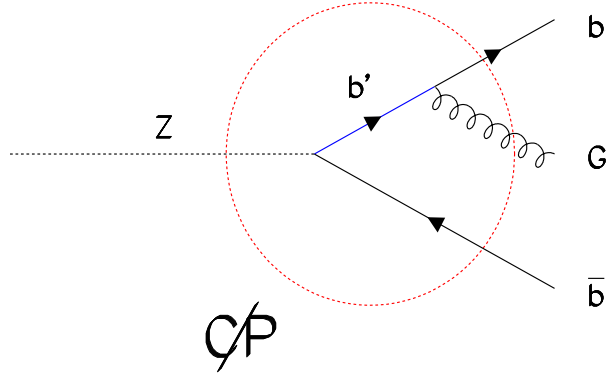}
    \parbox{0.9\textwidth}
{\caption{\it{Contribution to $Zb\bar{b}G$ from an excited quark $b'$. The
    diagram with the role of $b$ and $\bar{b}$ exchanged is not shown.
    }}\label{fig:excquark}
}  
  \end{center}
\end{figure}
%%%%%%%%%%%%%%%%%%%%%%%%%%%%%%end%%figure%%%%%%%%%%%%%%%%%%%%%%%%%%%%%

In this chapter we discuss the possible generation of 
chirality conserving CP-violating interactions as introduced in the
previous chapter in models with excited quarks.
Excitations of quarks would be natural in a
scenario where quarks have substructure and participate in a new type of
strong interaction. 
This type of models and effects from excited quarks at hadron colliders
have for instance been discussed in \cite{excq}.
In particular, we assume here that $b$ quarks have excited
partners $b'$, which could have spin $\frac{1}{2}$ or $\frac{3}{2}$. For
simplicity we consider a $b'$ of spin $\frac{1}{2}$ and mass $m_{b'}$. Due
to higher order dimensional operators in composite models
chirality-conserving $Zb'b$ couplings at the scale of GigaZ energies are a
priori possible (see e.g. \cite{Han:1998yr}). Because of
colour gauge invariance we expect the $b' b G$ couplings to be
chirality-flipping dipole couplings.
Then, couplings $\widehat{h}_{Vb,Ab}$ as
introduced in (\ref{lcp}) can be generated by the following
effective 
interactions of $b'$ and $b$ quarks, $Z$ bosons and gluons:
\begin{eqnarray}
  \label{lexcq} \nonumber
  {\cal L}'(x) = 
  & - & \frac{e}{2 \: \sin \vartheta_W\: \cos \vartheta_W}\; 
        Z_{\mu}(x)\: \bar{b}'(x)\: \gamma^{\mu} \; (g_V' - g_A' \gamma_5)\;
        b(x)\\ 
  & - & i \;\frac{g_s}{2 m_{b'}}\; \hat{d}_c \: \bar{b}'(x)\: \sigma^{\mu\nu}\:
  \gamma_5\: T^a\: b(x)\: G^a_{\mu\nu}(x) + {\rm h.c.}
\end{eqnarray}
Here $g_V'$, $g_A'$ and $\hat{d}_c$ are complex parameters, which can be
expected to be of order one if the underlying dynamics is strongly
interacting. In addition to $\hat{d}_c$, the chromoelectric dipole
transition form factor $b \rightarrow b'$, there will be in general also a
chromomagnetic transition form factor $\hat{d}_m$ which is omitted here for
brevity. 

The couplings $\widehat{h}_{Vb,Ab}$ have been calculated \cite{higgs} in
this
model from the diagrams of the type shown in Fig.~\ref{fig:excquark} for
$m_{b'} \gg m_Z$: 
\begin{eqnarray}
  \label{hexcf} \nonumber
  \widehat{h}_{Vb} & = & \frac{m_Z^2}{m_{b'}^2} \; {\rm Re}(\hat{d}_c \:
  g_A'^*) \;\; , \\
  \widehat{h}_{Ab} & = & - \frac{m_Z^2}{m_{b'}^2} \; {\rm Re}(\hat{d}_c \:
  g_V'^*) \;\; .
\end{eqnarray}

%%%%%%%%%%%%%%%%%%%%%%%%%%%%%%%%%%%%%%%%%%%%%%%%%%%%%%%%%%%%%%%%%%%%%%%%
%%%%%%%%%%%%%%%%%%%%%%%%%%%%%%%%%%%%%%%%%%%%%%%%%%%%%%%%%%%%%%%%%%%%%%%%
\section{Study of CP-violating couplings}
\label{sec:partons}
%%%%%%%%%%%%%%%%%%%%%%%%%%%%%%%%%%%%%%%%%%%%%%%%%%%%%%%%%%%%%%%%%%%%%%%%
%%%%%%%%%%%%%%%%%%%%%%%%%%%%%%%%%%%%%%%%%%%%%%%%%%%%%%%%%%%%%%%%%%%%%%%%

In our study we assume that one is able to
flavor-tag the $b$ quarks and to measure their momenta. This is justified due
to the extremely
good $b$-tagging capabilities foreseen at TESLA \cite{tesla}. For instance, the impact
parameter resolution at TESLA is expected to be about a factor 10 better
than at LEP \cite{moenig}.

The definition of a 3 and 4 jet sample requires the introduction of
resolution cuts. We use JADE cuts \cite{jade} requiring
\begin{equation}
        y_{ij} = \frac{2\, E_iE_j\,(1-\cos \vartheta_{ij})}{m_Z^2} >
        y_{cut} \;,
\label{jade}
\end{equation}
with $\vartheta_{ij}$ the angle between the momentum directions of any two
partons ($i \neq j$) and $E_i$, $E_j$ their energies in the Z rest
system.

%%%%%%%%%%%%%%%%%%%%%%%%%%%%%%%%%%%%%%%%%%%%%%%%%%%%%%%%%%%%%%%%%%%%%%%%
\subsection{CP-odd tensor and vector observables}
%%%%%%%%%%%%%%%%%%%%%%%%%%%%%%%%%%%%%%%%%%%%%%%%%%%%%%%%%%%%%%%%%%%%%%%%

We study our CP-violating couplings using
CP-odd observables constructed from the
momentum directions of the $b$ and $\bar{b}$ quarks,
$\widehat{\bf k}_b={\bf k}_b/|{\bf k}_b|$  and $\widehat{\bf k}_{\bar{b}}={\bf k}_{\bar{b}}/|{\bf k}_{\bar{b}}|$
(see \cite{zdecay,xsec,bernnach2,hab}):
\begin{equation}
        T_{ij} = (\widehat{\bf k}_{\bar{b}} - \widehat{\bf k}_b)_i \; (\widehat{\bf k}_{\bar{b}} \times
\widehat{\bf k}_b)_j \; + (i \leftrightarrow j) \; ,
\label{ten}
\end{equation}
\vspace*{-0.5cm}
\begin{equation}
        V_i = (\widehat{\bf k}_{\bar{b}} \times \widehat{\bf k}_b)_i \;  ,
\label{vec}
\end{equation}
with $i$, $j$ the Cartesian vector indices in the Z rest system.
 
The observables $T_{ij}$ transform as tensor components, $V_i$ as
vector components. For polarized $e^+e^-$ beams and our rotationally invariant cuts
(\ref{jade}) 
their expectation values are then proportional to the Z
tensor polarization $S_{ij}$ and vector polarization $s_i$,
respectively. For all definitions concerning the $Z$ density matrix see
section 2.1 of \cite{zdecay}.
Defining the positive $z$-axis in the $e^+$ beam
direction, we have
\begin{equation}
\label{s}
  {\bf s} = \left( \begin{array}{c}
  0 \\ 0 \\ s_3
\end{array} \right) \;,
\end{equation}
\begin{equation}
\label{sij}
  (S_{ij}) = \frac{1}{6} \left( \begin{array}{ccc}
  -1 & 0 & 0 \\
  0 &-1 & 0 \\
  0 & 0 & 2    
\end{array} \right) \;,
\end{equation}
where 
\begin{equation}
\label{s3}
        s_3 = \frac{ s_3^{(0)} (1 - P_+ P_-) + (P_+ - P_-)}{ (1 - P_+ P_-) + s_3^{(0)} (P_+ - P_-)}          
\end{equation}
and
\begin{equation}
\label{s30}
        s_3^{(0)} = \frac{2 \, g_{Ve} g_{Ae}}{g_{Ve}^2 + g_{Ae}^2} = 0.209\;,
\end{equation}
with $g_{Ve}=-1/2+2 \sin^2 \vartheta_W$ and $g_{Ae}=-1/2$ the weak vector
and axial vector $Zee$
couplings. $P_+$ and $P_-$ are the longitudinal polarizations for
positron and electron, respectively, measured in the direction of the
particle's velocity. We have $|P_\pm| \le 1$. From (\ref{s}) ---
(\ref{s30}) we see that the components $T_{33}$ and $V_3$
are the most sensitive ones.

Note that the tensor observables do {\em not} change their sign upon charge
misidentification ($\widehat{\bf k}_{\bar{b}} \leftrightarrow \widehat{\bf k}_b$) whereas the
vector observables do. 
Thus, it is only for the measurement of the latter
that charge identification is indispensable.
%, which makes the vector
%observables less valuable for the experimental analysis.

We have computed the expectation values of the observables (\ref{ten}),
(\ref{vec}) for different JADE cuts (\ref{jade}), as function of
\begin{equation}
\label{hat}
        \widehat{h}_b = \widehat{h}_{Ab} g_{Vb} - \widehat{h}_{Vb} g_{Ab}
\end{equation}
and 
\begin{equation}
\label{tilde}
        \widetilde{h}_b = \widehat{h}_{Vb} g_{Vb} - \widehat{h}_{Ab} g_{Ab} \;,
\end{equation}
where
\begin{equation}
  g_{Vb} = -\frac{1}{2} + \frac{2}{3} \sin^2 \vartheta_W \;, \;\; g_{Ab}=
  -\frac{1}{2} \;.
\end{equation}

The expectation value of a CP-odd observable ${\cal{O}}$
has the following general form:
\begin{equation}
        <\!{\cal{O}}\!> = (c_1 \widehat{h}_b + c_2 \widetilde{h}_b) \,
        \frac{\Gamma^{SM}_{3/4\;jets}}{\Gamma_{3/4\;jets}} \;.
\label{cpoddobs}
\end{equation}
Here $c_{1,2}$ are constants, $\Gamma^{SM}_{3/4\;jets}$ and
$\Gamma_{3/4\;jets}$ denote the 
corresponding \ZTJ\ and \ZJ\
decay widths in the SM and in the theory with SM plus CP-violating couplings,
respectively. Note that terms quadratic in the anomalous couplings
  are CP-even. Thus on the r.h.s. of (\ref{cpoddobs}) they only emerge in
  $\Gamma_{3/4\;jets}$.
In an experimental analysis 
one has two options. The first one is to study directly the expectation
values $<\!{\cal{O}}\!>$ which have a non-linear dependence on \hb, \hbn.
This dependence becomes linear for small anomalous couplings. In the
following we 
will neglect non-linear terms, that is assume
$\Gamma_{3/4\;jets} \approx \Gamma^{SM}_{3/4\;jets}$. The other option is
to take
$\Gamma^{SM}_{3/4\;jets}$ from the theoretical calculation,
$\Gamma_{3/4\;jets}$ and 
$<\!{\cal{O}}\!>$ from the experimental measurement. The quantity
$<\!{\cal{O}}\!> \! \cdot \,\Gamma_{3/4\;jets}$ is then an observable strictly linear in
the anomalous couplings,
which has obvious advantages.

For unpolarised $e^+ e^-$ beams a non-zero value $<\!{\cal{O}}\!> \, \neq 0$
for one of our CP-odd observables above is an unambiguous indicator of CP
violation. For longitudinally polarised beams this holds if possible
chirality flipping interactions at the $e^+e^- Z$ vertex --- which do not
exist in the SM --- are neglected. See~\cite{dnn} for an extensive
discussion of this point.

From the measurement of a single observable (\ref{cpoddobs}) we can get a
simple estimate of its sensitivity to \hb\ by assuming $\hbn=0$. The
error on a measurement of \hb\ is then to leading order in the anomalous
couplings:
\begin{equation}
        \delta\widehat{h}_b =
        \frac{\sqrt{<\!{\cal{O}}^2\!>_{SM}}}{|c_1| \sqrt{N}} \;,
\label{dhb}
\end{equation}
where $N$ is the number of events within cuts. Similarly, assuming
$\hb=0$ we get the error on \hbn\ as 
\begin{equation}
        \delta\widetilde{h}_b =
        \frac{\sqrt{<\!{\cal{O}}^2\!>_{SM}}}{|c_2| \sqrt{N}} \;.
\label{dhbn}
\end{equation}
A measure for the sensitivity of \OBS\ to \hb\ (\hbn) is then $1/\dhb$
($1/\dhbn$).

In very good
approximation, it was found for \ZTJ\ and \ZJ\ that the tensor observables
are only sensitive to 
$\widehat{h}_b$ and the vector observables only to 
$\widetilde{h}_b$.
A detailed discussion about that can be found in
\cite{width,hab,4jet,dok}. 

A measurement of \hb\ , \hbn\ has to produce a mean
value larger than $\delta\widehat{h}_b$ (\ref{dhb}),
$\delta\widetilde{h}_b$ (\ref{dhbn})
to be able to claim a
non-zero effect at the 1 s.~d. level. 

%%%%%%%%%%%%%%%%%%%%%%%%%%%%%%%%%%%%%%%%%%%%%%%%%%%%%%%%%%%%%%%%%%%%%%%%
\subsection{Numerical results}
%%%%%%%%%%%%%%%%%%%%%%%%%%%%%%%%%%%%%%%%%%%%%%%%%%%%%%%%%%%%%%%%%%%%%%%%
\label{sec:partres}

We have calculated the sensitivities to
$\widehat{h}_b$ and $\widetilde{h}_b$ 
for the tensor (\ref{ten}) and vector (\ref{vec})
observables varying the jet resolution parameter \ycut. 
Comparing with optimal observables
it was found for unpolarized
beams \cite{hab,4jet} that these simple observables (\ref{ten},\ref{vec})
reach nearly optimal sensitivities. 
%This remains true for longitudinal polarization of the beam, because
%the Z-production process is only modified by factors. 
Therefore optimal
observables are not considered in the following.

We assume a total number of $N_{tot} = 10^9$ Z decays for unpolarized beams,
following the GigaZ scenario. The number $N$ of events within cuts which is
available for the 
analysis is then given by
\begin{equation}
  \label{nevents}
     N_{3/4\;jets} = N_{tot} \frac{\Gamma^{SM}_{3/4\;jets}}{\Gamma_Z} \;,
\end{equation}
with ${\Gamma_Z}$ being the total Z decay width.
Solely due to higher statistics in
GigaZ of about a
factor of one hundred compared to the sum of the four LEP experiments, the
sensitivity to the CP violating 
couplings increases by a factor 10, as can be seen from (\ref{dhb},
\ref{dhbn}). 

\begin{figure}[H]
\begin{center}
\begin{picture}(0,0)
\put(-230,65){\epsfig{file=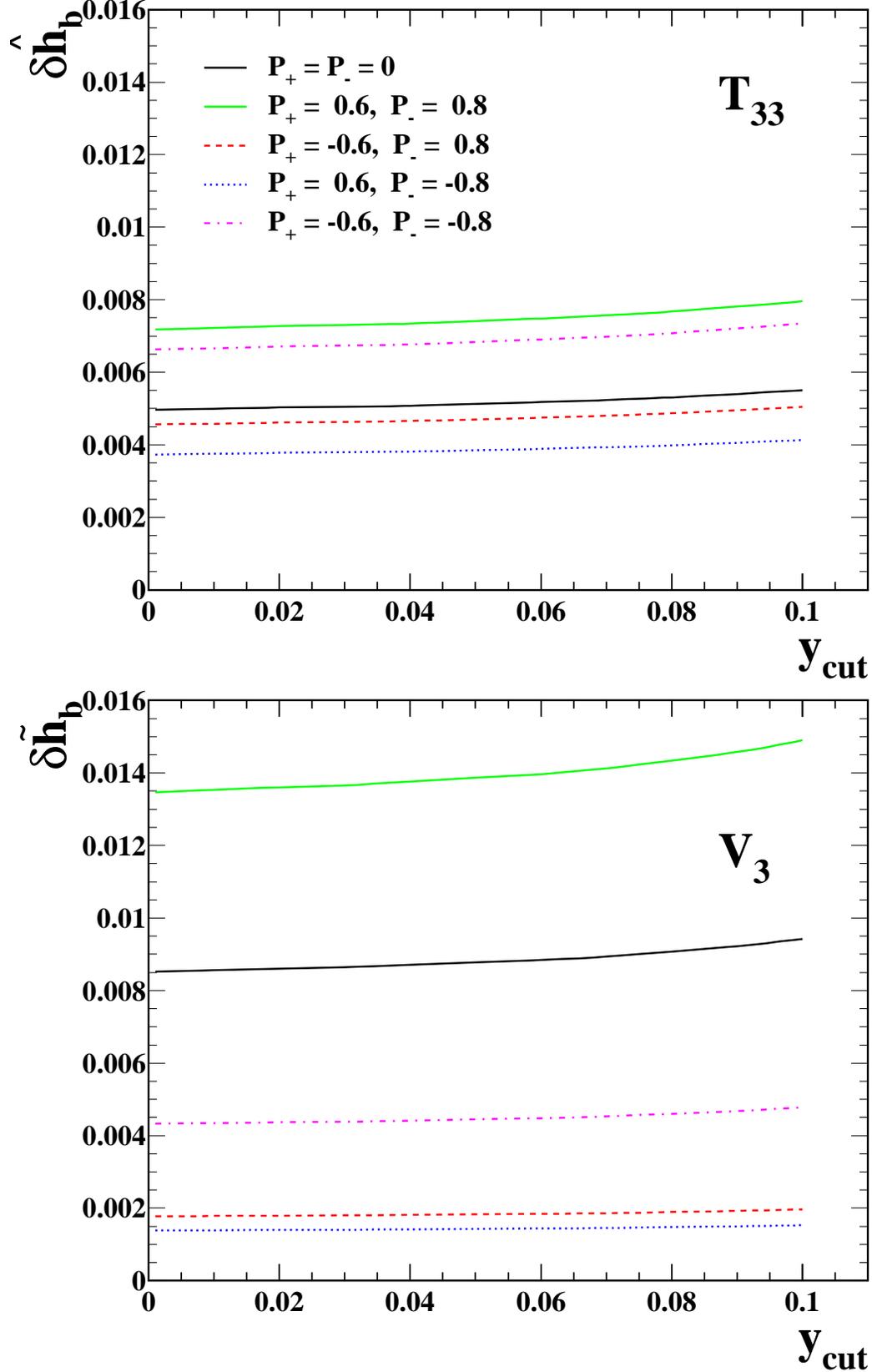,width=0.85\hsize,bbllx=35,bblly=760,bburx=500,bbury=35,clip=}} 
\end{picture}
\vspace{19.7cm}
{\caption{\it{
The inverse sensitivities of tensor $T_{33}$ and vector $V_3$ observables to
$\widehat{h}_b$ and $\widetilde{h}_b$ (\ref{hat},\ref{tilde}) obtainable in \ZTJ,
as function of the jet resolution parameter $y_{cut}$ (\ref{jade}) for
different longitudinal polarizations of the $e^+$ and $e^-$ beams assuming
an integrated luminosity which would lead to $10^9$ Z decays without
polarization.
    }}\label{fig:del_asym_3jets}
}  \end{center}
\end{figure}

\begin{figure}[H]
\begin{center} 
\begin{picture}(0,0)
\put(-230,65){\epsfig{file=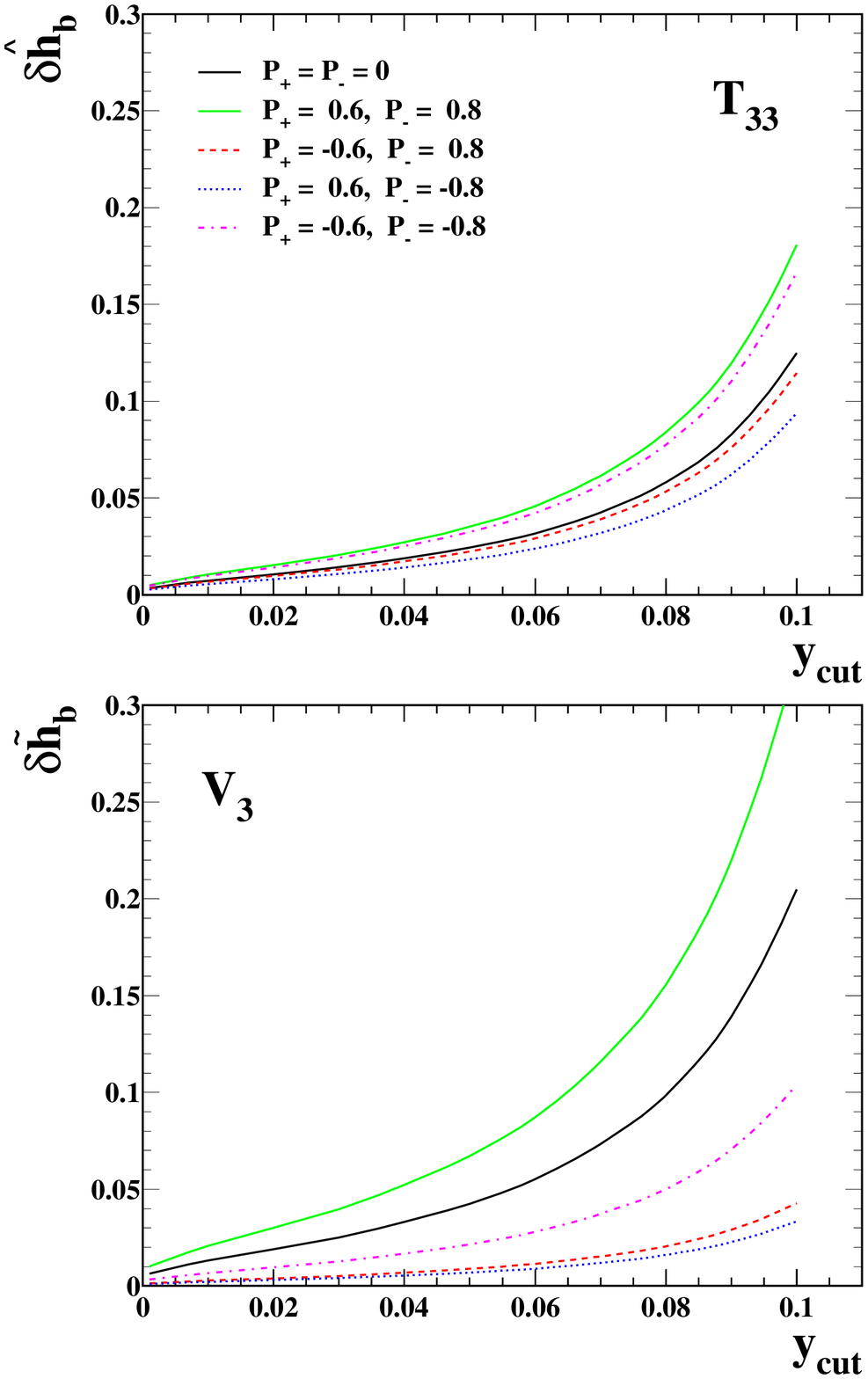,width=0.85\hsize,bbllx=35,bblly=760,bburx=500,bbury=35,clip=}} 
\put(-140,-60){\epsfig{file=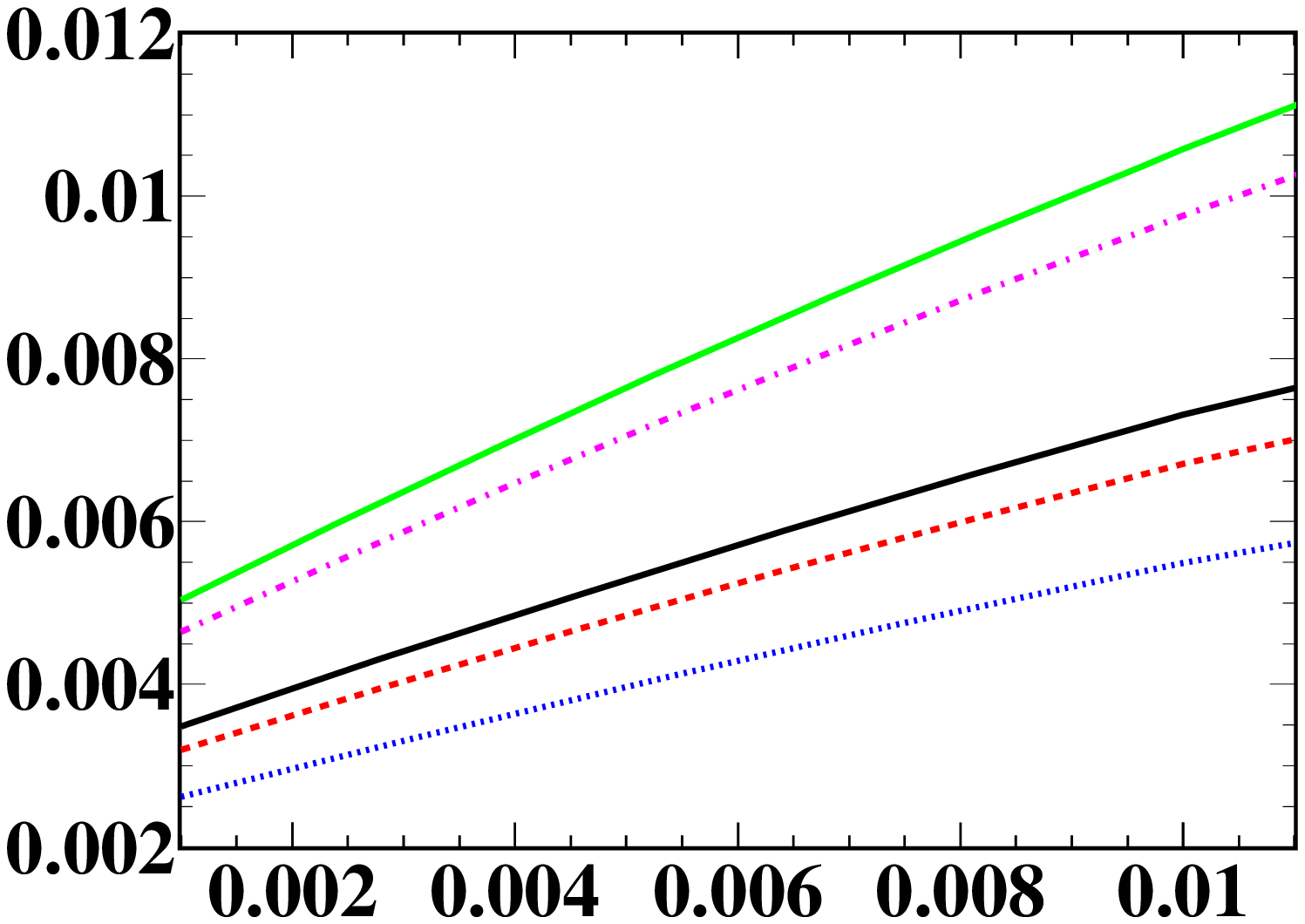,width=0.31\hsize,bbllx=65,bblly=730,bburx=505,bbury=405,clip=}}
\put(-140,-333){\epsfig{file=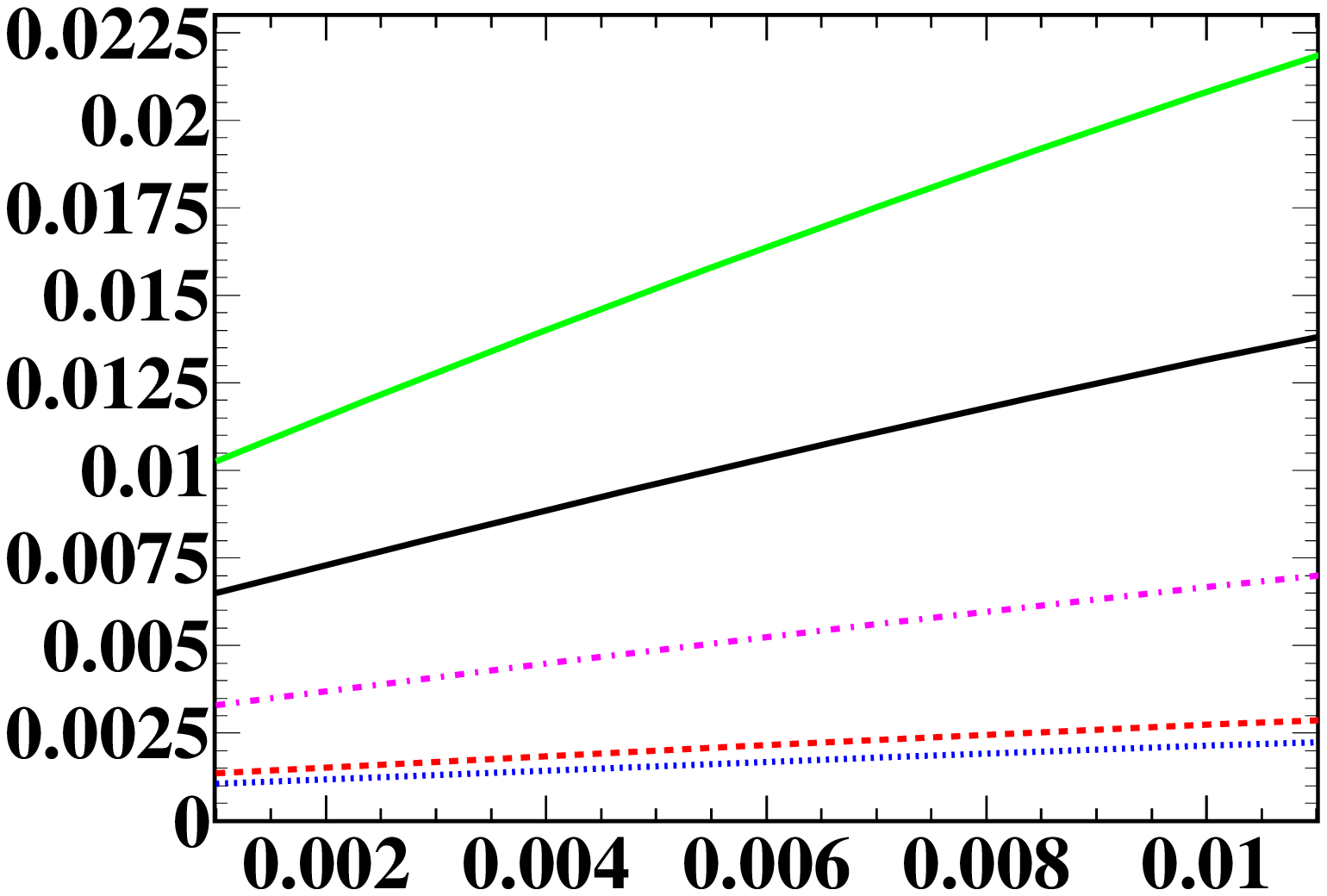,width=0.31\hsize,bbllx=65,bblly=730,bburx=505,bbury=405,clip=}}
\end{picture}
\vspace{19.7cm}
{\caption{\it{
The inverse sensitivities of tensor $T_{33}$ and vector $V_3$ observables to
$\widehat{h}_b$ and $\widetilde{h}_b$ (\ref{hat},\ref{tilde}) obtainable in \ZJ,
as function of the jet resolution parameter $y_{cut}$ (\ref{jade}) for
different longitudinal polarizations of the $e^+$ and $e^-$ beams assuming
an integrated luminosity which would lead to $10^9$ Z decays without
polarization.
    }}\label{fig:del_asym_4jets}
}  \end{center}
\end{figure}

The inverse sensitivities to these
CP-odd couplings as calculated from (\ref{dhb}) and (\ref{dhbn}),
respectively, are shown in Fig.~\ref{fig:del_asym_3jets} for \ZTJ\ and in
Fig.~\ref{fig:del_asym_4jets} for \ZJ\ for different longitudinal beam
polarizations. 
The sensitivity decreases with increasing $y_{cut}$ for all
observables due to the decrease in number of events available. 

Because the expectation value of the tensor observable does not depend on
longitudinal polarization (\ref{sij}), the differences in
\dhb\ for 
different polarization choices reflect
only the change in statistics. For
$P_+ = 0.6 $ and $P_- = -0.8$ the enhancement of the Z production rate is
largest. The differences in \dhbn\ reflect both the change in statistics
and the modification of the expectation value due to polarization
(\ref{s3}). For 
$P_+ = 0.6 $ and $P_- = -0.8$ the sensitivity increases by more than a
factor of six 
compared to unpolarized beams. A convenient choice of the polarizations can
even lead to a better sensitivity of the vector observable to \hbn\ than of
the tensor observable to \hb.

\begin{figure}[H]
\begin{center} 
\begin{picture}(0,0)
\put(-230,65){\epsfig{file=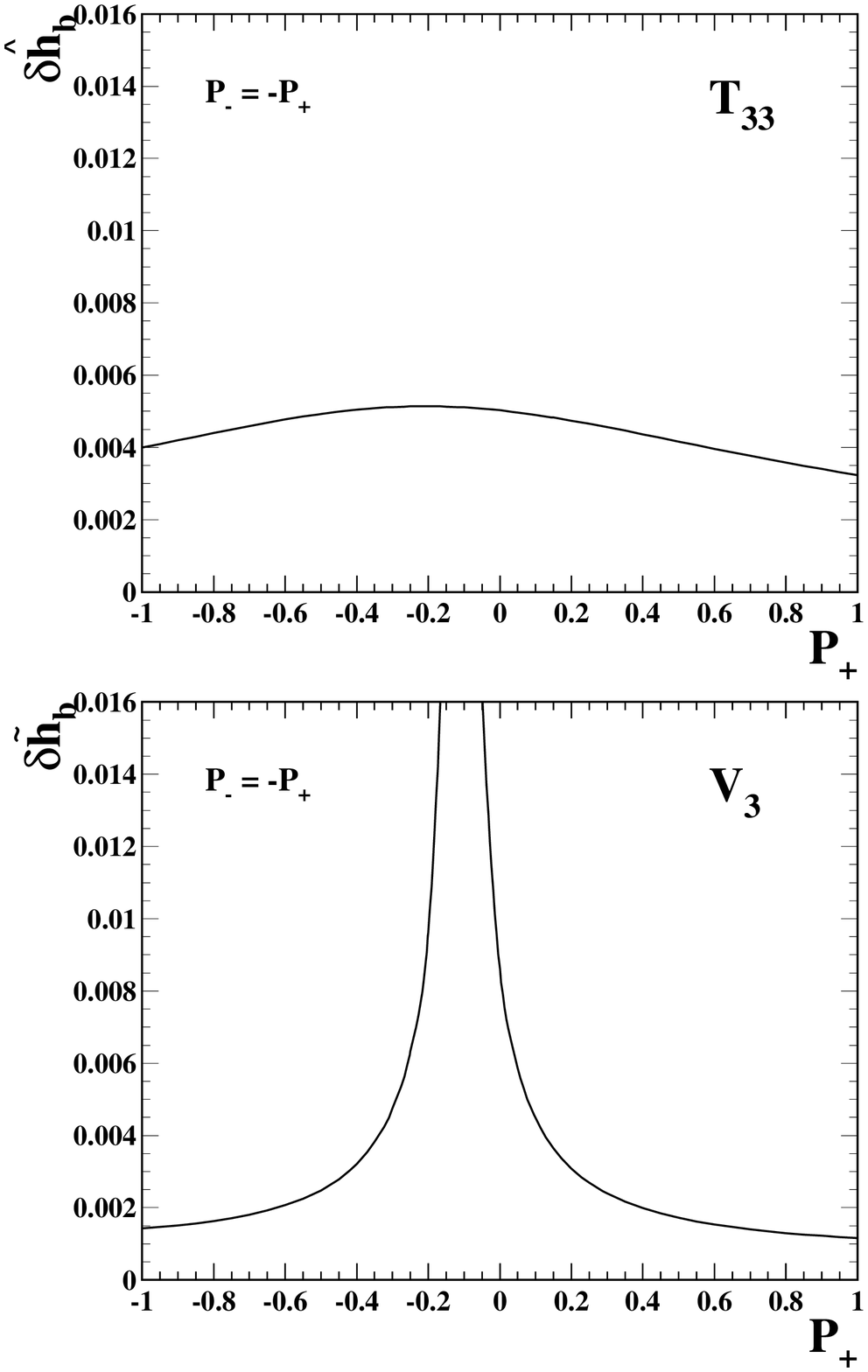,width=0.85\hsize,bbllx=35,bblly=760,bburx=500,bbury=35,clip=}} 
\end{picture}
\vspace{19.7cm}
{\caption{\it{
The inverse sensitivities of tensor $T_{33}$ and vector $V_3$ observables to
$\widehat{h}_b$ and $\widetilde{h}_b$ (\ref{hat},\ref{tilde}) obtainable in
\ZTJ\ for $y_{cut}=0.02$,
as function of the $e^+$ beam polarization for $P_- = -P_+$
assuming
an integrated luminosity which would lead to $10^9$ Z decays without
polarization.
    }}\label{fig:del_sym_3jets}
}  \end{center}
\end{figure}

\begin{figure}[H]
\begin{center}
\begin{picture}(0,0)
\put(-230,65){\epsfig{file=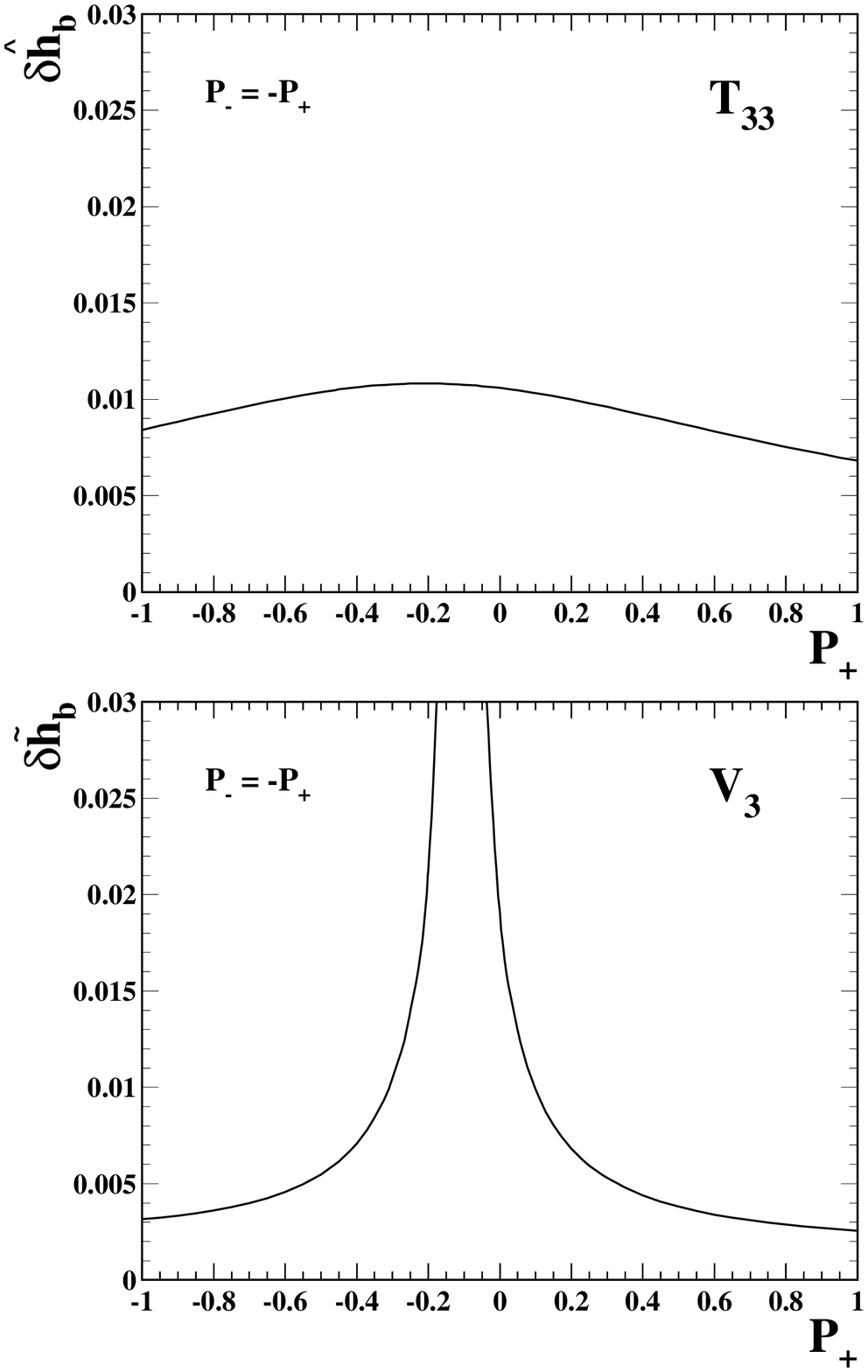,width=0.85\hsize,bbllx=35,bblly=760,bburx=500,bbury=35,clip=}} 
\end{picture}
\vspace{19.7cm}
{\caption{\it{
The inverse sensitivities of tensor $T_{33}$ and vector $V_3$ observables to
$\widehat{h}_b$ and $\widetilde{h}_b$ (\ref{hat},\ref{tilde}) obtainable in
\ZJ\ for $y_{cut}=0.02$,
as function of the $e^+$ beam polarization for $P_- = -P_+$
assuming
an integrated luminosity which would lead to $10^9$ Z decays without
polarization.
    }}\label{fig:del_sym_4jets}
}  \end{center}
\end{figure}

In contrast, an unsuitable choice of the polarizations could kill any
sensitivity of the vector observable. This is illustrated in
Figs.~\ref{fig:del_sym_3jets} and \ref{fig:del_sym_4jets} for \ZTJ\ and
\ZJ, respectively: The inverse sensitivities are shown as a function of
the positron polarization assuming $P_- = -P_+$. For $P_+ \simeq - 0.1$ the
expectation value for the vector observable and therefore the sensitivity
to \hbn\ vanishes. For the tensor observable this cannot happen because
the sensitivity to \hb\ depends on the polarization only due to the change
in the total number of Z decays. 

It should be stressed that in this article we present a tree-level
calculation. Thus next-to-leading order QCD corrections are not taken into
account. For the SM part 
they can be found in \cite{NLO3jets} for \ZTJ\ including non-vanishing
$b$ quark masses and in \cite{NLO4jets} for \ZJ\ for massless
quarks. QCD corrections to the anomalous couplings (\ref{lcp}) could be
calculated using 
\hphantom{see
 (\ref{cpoddobs}--\ref{dhbn}), these corrections can be expected}
\vspace*{-0.85cm}
\newpage
%%%%%%%%%%%%%%%%%%begin%%figure%%%%%%%%%%%%%%%%%%%%%%%%%%%%%%%%%%%%%%%
%%%%%%%%%%%%%%%%%%%%%%%%%%%%%%end%%figure%%%%%%%%%%%%%%%%%%%%%%%%%%%%%
%
%%%%%%%%%%%%%%%%%%begin%%figure%%%%%%%%%%%%%%%%%%%%%%%%%%%%%%%%%%%%%%%
%%%%%%%%%%%%%%%%%%%%%%%%%%%%%%end%%figure%%%%%%%%%%%%%%%%%%%%%%%%%%%%%
%
%%%%%%%%%%%%%%%%%%begin%%figure%%%%%%%%%%%%%%%%%%%%%%%%%%%%%%%%%%%%%%%
%%%%%%%%%%%%%%%%%%%%%%%%%%%%%%end%%figure%%%%%%%%%%%%%%%%%%%%%%%%%%%%%
%
%%%%%%%%%%%%%%%%%%begin%%figure%%%%%%%%%%%%%%%%%%%%%%%%%%%%%%%%%%%%%%%
%%%%%%%%%%%%%%%%%%%%%%%%%%%%%%end%%figure%%%%%%%%%%%%%%%%%%%%%%%%%%%%%
%
%%%%%%%%%%%%%%%%%%begin%%figure%%%%%%%%%%%%%%%%%%%%%%%%%%%%%%%%%%%%%%%
\begin{figure}[H]
\begin{center}
\begin{picture}(0,0)
\put(-230,65){\epsfig{file=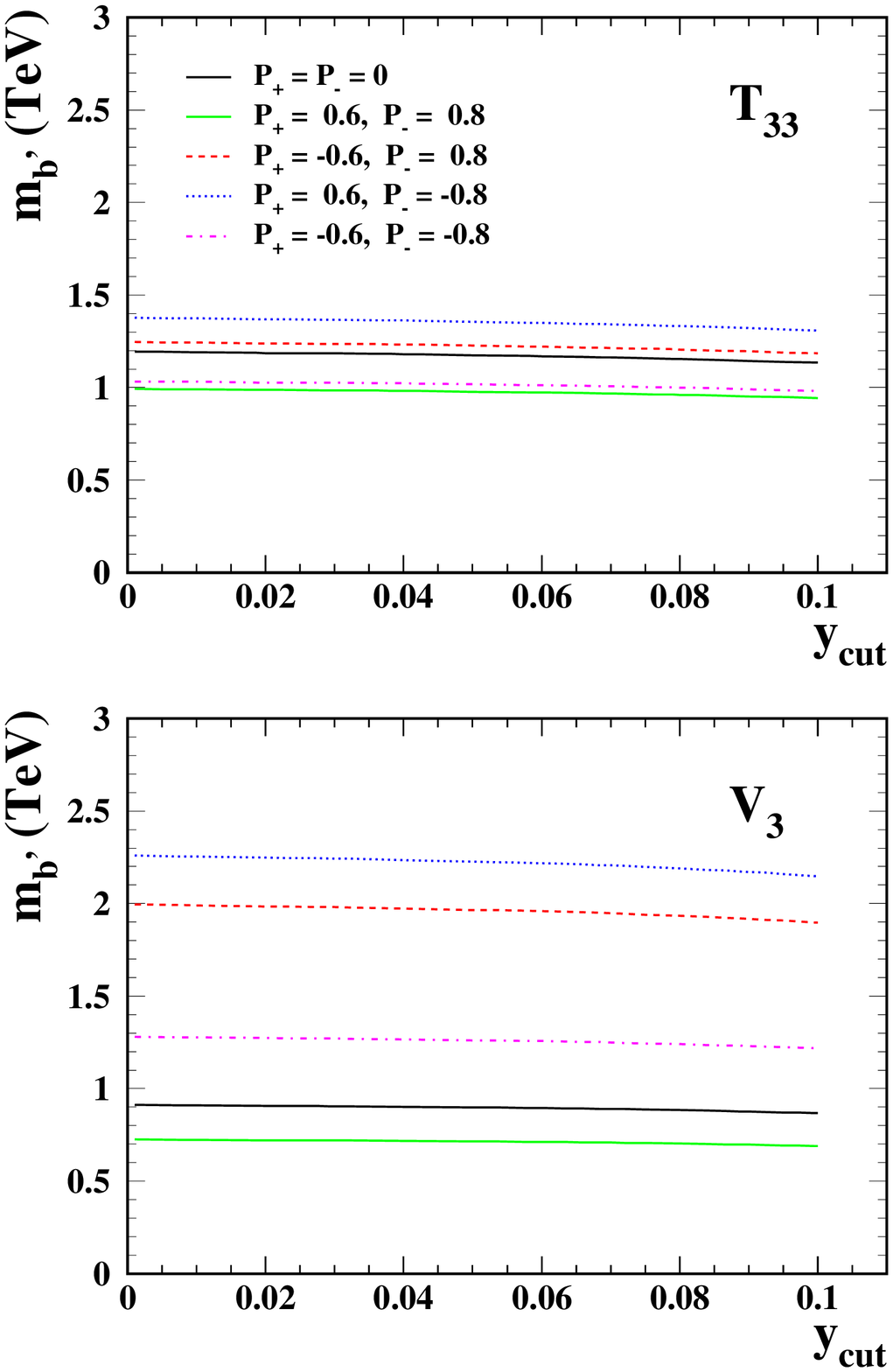,width=0.85\hsize,bbllx=35,bblly=760,bburx=500,bbury=35,clip=}} 
\end{picture}
\vspace{19.cm}
{\caption{\it{
Lower limits on the excited quark mass $m_{b'}$ at the 1 s.~d. level
which can be derived from a 
measurement of tensor $T_{33}$ and vector $V_3$ observables in \ZTJ,
as function of the jet resolution parameter $y_{cut}$ (\ref{jade}) for
different longitudinal polarizations of the $e^+$ and $e^-$ beams assuming
an integrated luminosity which would lead to $10^9$ Z decays without
polarization. Couplings for the $b'$ as discussed in the text are assumed.
    }}\label{fig:mb_asym_3jets}
}  \end{center}
\end{figure}
%%%%%%%%%%%%%%%%%%%%%%%%%%%%%%end%%figure%%%%%%%%%%%%%%%%%%%%%%%%%%%%%
%
%%%%%%%%%%%%%%%%%%begin%%figure%%%%%%%%%%%%%%%%%%%%%%%%%%%%%%%%%%%%%%%
\begin{figure}[H]
\begin{center} 
\begin{picture}(0,0)
\put(-230,65){\epsfig{file=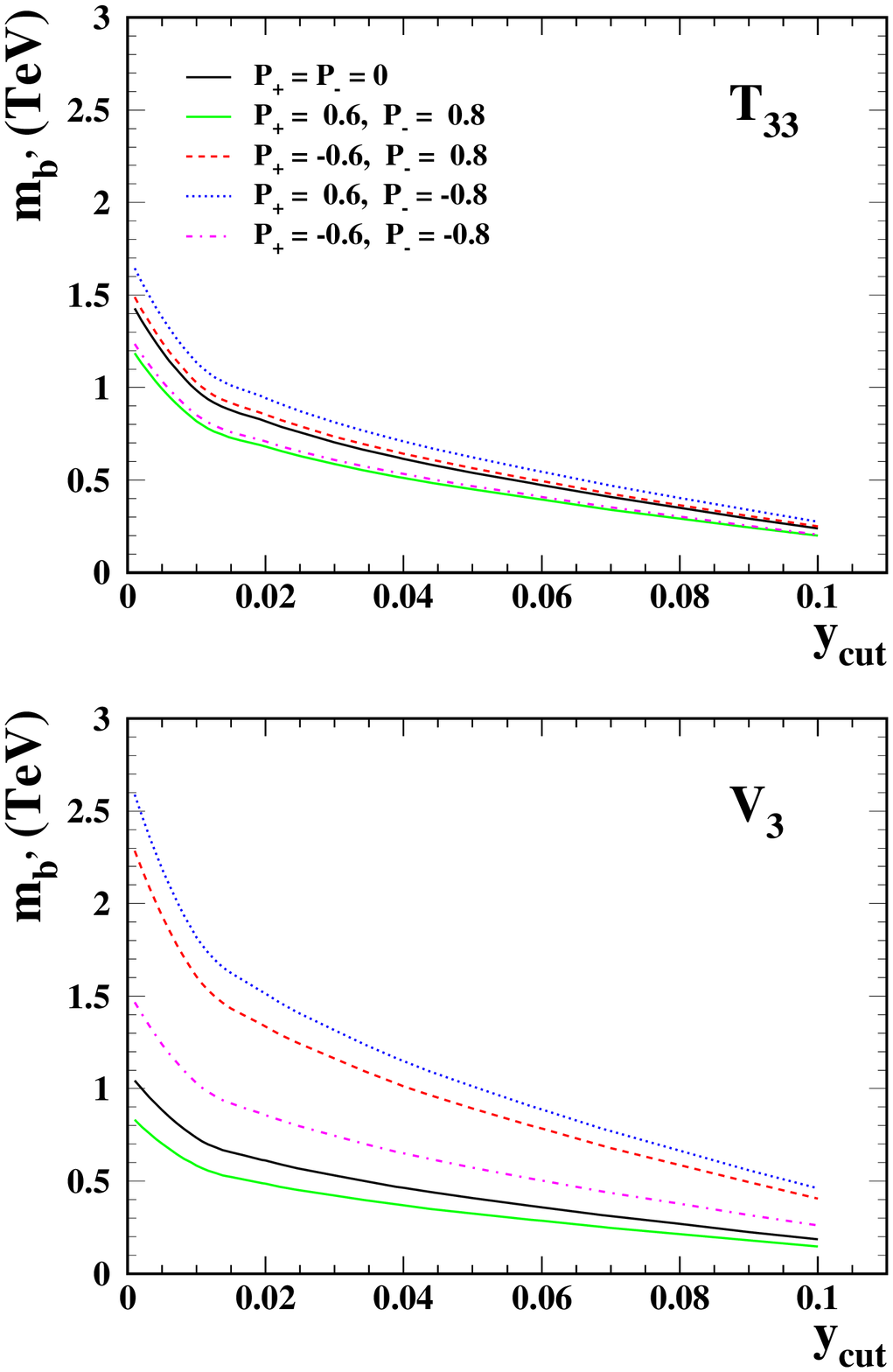,width=0.85\hsize,bbllx=35,bblly=760,bburx=500,bbury=35,clip=}} 
\end{picture}
\vspace{19.cm}
{\caption{\it{
Lower limits on the excited quark mass $m_{b'}$ at the 1 s.~d. level
which can be derived from a
measurement of tensor $T_{33}$ and vector $V_3$ observables in \ZJ,
as function of the jet resolution parameter $y_{cut}$ (\ref{jade}) for
different longitudinal polarizations of the $e^+$ and $e^-$ beams assuming
an integrated luminosity which would lead to $10^9$ Z decays without
polarization. Couplings for the $b'$ as discussed in the text are assumed.
    }}\label{fig:mb_asym_4jets}
}  \end{center}
\end{figure}
%%%%%%%%%%%%%%%%%%%%%%%%%%%%%%end%%figure%%%%%%%%%%%%%%%%%%%%%%%%%%%%%
%
\newpage
\hspace{-0.79cm} the methods of effective field theories (see for instance
\cite{weinberg}). However, 
because here we always consider ratios of expectation values, see
(\ref{cpoddobs}--\ref{dhbn}), these corrections can be expected to cancel
to some extent and to lead 
only to moderate changes to the numbers given. This should hold at least
for not too low values of $\ycut$. From \cite{NLO3jets,NLO4jets} one finds
higher order QCD corrections to become important for $\ycut \, \ltap \, 0.01$.
Thus, to be on the safe side one should restrict the analysis to $\ycut\, \gtap
\, 0.01$.

%%%%%%%%%%%%%%%%%%%%%%%%%%%%%%%%%%%%%%%%%%%%%%%%%%%%%%%%%%%%%%%%%%%%%%%%
\subsection{Interpretation in the framework of excited quarks}
%%%%%%%%%%%%%%%%%%%%%%%%%%%%%%%%%%%%%%%%%%%%%%%%%%%%%%%%%%%%%%%%%%%%%%%%

If a measurement of \hb\ , \hbn\ produces a mean
value lower than $\delta\widehat{h}_b$ (\ref{dhb}),
$\delta\widetilde{h}_b$ (\ref{dhbn})
a non-zero effect at the 1 s.~d. level cannot be claimed and therefore an
upper limit on these couplings can be derived. Using (\ref{hexcf}) this can
be translated into lower bounds on the excited quark mass
$m_{b'}$. Assuming ${\rm Re}(\hat{d}_c \: g_A'^*) = {\rm Re}(\hat{d}_c \:
g_V'^*) = 1 $ these bounds are shown in Fig.~\ref{fig:mb_asym_3jets} for
\ZTJ\ and in
Fig.~\ref{fig:mb_asym_4jets} for \ZJ\ for different longitudinal beam
polarizations. 

In \cite{excq_cdf} at the $95\%$ confidence level excited quarks with mass
between 80 and 570 GeV and between 580 and 760 GeV were excluded. In
\cite{excq_d0} the lower limit $m_{q'} > 775$~GeV on the masses of
excited quarks was given.\footnote{These numbers should be compared to our
  excited quark mass limits at the 2 s. d. level. In that case a
  measurement of \hb\ , \hbn\ has to produce a mean value larger than
  $2\, \delta\widehat{h}_b$ (\ref{dhb}), $2\, \delta\widetilde{h}_b$
  (\ref{dhbn}) 
  to be able to claim a non-zero effect. From (\ref{hexcf}) one derives
  that the mass limits at the 1 s. d. level given in
  Fig.~\ref{fig:mb_asym_3jets} and~\ref{fig:mb_asym_4jets} have to be
  divided by a factor $\sqrt{2}$ to get the limits at the 2 s. d. level.} 
However, these results apply to excited $u$ and $d$
quarks only and do not exclude a lighter $b'$ quark.

%%%%%%%%%%%%%%%%%%%%%%%%%%%%%%%%%%%%%%%%%%%%%%%%%%%%%%%%%%%%%%%%%%%%%%%%
%%%%%%%%%%%%%%%%%%%%%%%%%%%%%%%%%%%%%%%%%%%%%%%%%%%%%%%%%%%%%%%%%%%%%%%%
\section{Conclusions}
%%%%%%%%%%%%%%%%%%%%%%%%%%%%%%%%%%%%%%%%%%%%%%%%%%%%%%%%%%%%%%%%%%%%%%%%
%%%%%%%%%%%%%%%%%%%%%%%%%%%%%%%%%%%%%%%%%%%%%%%%%%%%%%%%%%%%%%%%%%%%%%%%

In this paper, we have reviewed calculations concerning the search
for CP violation in the 3 jet and 4 jet decays of the Z boson with at least
two of 
the jets 
originating from $b$ and $\bar{b}$ quarks. 
We have studied a CP-violating
contact interaction with a vector and axial vector coupling \hvb, \hab\
(\ref{lcp}), (\ref{hhat}). We have discussed how such
couplings can be generated in models with an excited $b$ quark, $b'$.
Such couplings can also arise at one loop level in multi-Higgs
extensions of the Standard Model \cite{higgs,bernnachhiggs}. 
Longitudinal beam polarization is included.

We studied a tensor and vector observable 
which can be used for the measurement of the
anomalous couplings.
While the sensitivity of the tensor observable to CP-violating effects is only
affected by the variation of statistics due to beam polarization given a
certain integrated luminosity, the
expectation value of the vector observable itself changes by the factor
$s_3$ (\ref{s3}).

If flavor tagging of  $b$ and
\bb\ jets is available then, with a total number of
$10^9$ Z decays and choosing a cut
parameter\footnote{This value of \ycut\ is, in fact, a relatively large
  number for a selection of events \ZJ. So the numbers given in the
  following are 
  conservative for this channel.}
$y_{cut} = 0.02$, the anomalous coupling constant
\hb\ (\ref{hat}) can be determined with an accuracy of order 0.004 (\ZTJ)
and 0.008 (\ZJ) at 1~s.~d. level using the tensor observable $T_{33}$
(\ref{ten}) for the measurement. Here, $b - \bb$ distinction is not necessary.
These accuracies are close to the ones which already can be obtained with
unpolarized beams. If in a measurement a non-zero effect at the 1
s.~d. level is not observed excited quark masses $m_{b'}$ lower than 1.4 TeV
(\ZTJ) and 0.94 TeV (\ZJ) can be excluded if appropriate
couplings are of a size characteristic of a strong interaction.

If $b - \bb$ distinction is experimentally realizable, which should be the
case at a future linear collider, the coupling constant
\hbn\ (\ref{tilde}) can be measured with an accuracy of order 0.0015 (\ZTJ) and
0.003 
(\ZJ) using the vector observable 
$V_{3}$ (\ref{vec}) and choosing $P_+ = 0.6 $ and $P_- = -0.8$ as
longitudinal polarizations of positron and electron, respectively.
In case of a non-observation of an effect at the 1 s.~d. level excited
quark masses $m_{b'}$ lower than 2.2 TeV (\ZTJ) and 1.5 TeV (\ZJ) can be
excluded if the relevant
couplings are of a size characteristic of a strong interaction. 

Comparing 3 and 4 jet analyses we found that the sensitivity to the
anomalous coupling \hb\ was roughly constant as function of the cut
parameter \ycut\ for $\ycut<0.1$ in the 3 jet case. For the 4 jet case the
sensitivity was found to increase as \ycut\ decreases. For $\ycut \approx 0.01$
the 4 jet sensitivity was found to become equal to that from 3
jets.
Of course in an
experimental analysis one should try to make both 3 and 4 jet analyses in
order to extract the maximal possible information from the data.

In our theoretical investigations we assumed always $100\%$ efficiencies
and considered the statistical errors only. 
Assuming systematic errors to be of the same size as the statistical ones, the accuracies in the determinations of \hb, \hbn\ discussed
above should indeed be better by more than one order of magnitude than
those derived from LEP.
As shown in \cite{higgs,bernnachhiggs} this will, for instance, give
valuable information 
on the scalar sector in multi-Higgs extensions of the Standard Model. That
interesting information on models with excited quarks can be derived as well
has been discussed in detail here.

\enlargethispage{\baselineskip}
%%%%%%%%%%%%%%%%%%%%%%%%%%%%%%%%%%%%%%%%%%%%%%%%%%%%%%%%%%%%%%%%%%%%%%%%
\subsection*{Acknowledgements}
%%%%%%%%%%%%%%%%%%%%%%%%%%%%%%%%%%%%%%%%%%%%%%%%%%%%%%%%%%%%%%%%%%%%%%%%
We would like to thank A.~Brandenburg, F.~Nagel, T.~Ohl, S.~Schmitt and
M.~Spira for valuable discussions. We also thank A.~Brandenburg, A.~Denner,
G.~Pasztor and M.~Spira for carefully reading the manuscript.

%\newpage

%%%%%%%%%%%%%%%%%%%%%%%%%%%%%%%%%%%%%%%%%%%%%%%%%%%%%%%%%%%%%%%%%%%%%%%%
%%% References
%%%%%%%%%%%%%%%%%%%%%%%%%%%%%%%%%%%%%%%%%%%%%%%%%%%%%%%%%%%%%%%%%%%%%%%%
%\baselineskip15pt
%\begin{thebibliography}{19}
%\newpage

\thispagestyle{plain}
\pagestyle{empty}
\end{document}